\documentclass[useAMS,usenatbib,usegraphicx]{mn2e}
\usepackage{psfig}   
\usepackage{graphicx}
\usepackage{subfigure}

\title[The role of cluster evolution in disrupting planetary systems: the Kozai mechanism]{The role of cluster evolution in disrupting planetary systems and disks: the Kozai mechanism}

\author[R.~J.~Parker and  S.~P.~Goodwin]{
  Richard J.~Parker\thanks{E-mail: r.parker@sheffield.ac.uk} and Simon P.~Goodwin\\
   Department of Physics and Astronomy, University of Sheffield, Sheffield, S3 7RH, UK}

\begin{document}

\date{}
                             
\pagerange{\pageref{firstpage}--\pageref{lastpage}} \pubyear{2009}

\maketitle

\label{firstpage}

\begin{abstract}
We examine the effects of dynamical evolution in clusters on planetary systems or 
protoplanetary disks orbiting the components of binary stars. In particular, we look for evidence 
that the companions of host stars of planetary systems or disks could have their inclination angles raised from 
zero to between the threshold angles (39.23$^\circ$ and 140.77$^\circ$) that can induce the Kozai mechanism. 
We find that up to 20 per cent of binary systems have their inclination angles increased to within the threshold range. 
Given that half of all extrasolar planets could be in binary systems, we suggest that up to 10 per cent of 
extrasolar planets could be affected by this mechanism.
\end{abstract}


\section{Introduction}

The majority of star formation is believed to occur in clustered environments 
\citep[][and references therein]{Lada03}. Furthermore, around 60 per cent of 
solar-type stars are observed in binary systems in the field 
\citep{Duquennoy91}. It is believed that a higher percentage of these stars 
(and probably \emph{all} of them) form not as singles, but in binary or higher 
order systems \citep[e.g.][]{Goodwin05,Goodwin07}, although see \citet{Lada06} 
for an alternative interpretation.

Several studies \citep[e.g.][]{Kroupa95a,Kroupa95b,Parker09a} have 
demonstrated that intense dynamical evolution occurs for stellar systems 
in typical (i.e. Orion-like) clusters. If most stars form in such clusters, 
then this dynamical processing may also have a significant effect on any planet 
formation that occurs around binary stars.

Over 300 extrasolar planets have been discovered to date\footnote{The number of discoveries is increasing 
on an almost daily basis; for the most up-to-date catalogue maintained by J. Schneider 
see http://exoplanet.eu/.}. Of the nearby extrasolar planets 
\citep[i.e. within 200\,pc, see][]{Butler06}\footnote{See  http://exoplanets.org/planets.shtml 
for the census by \citet{Butler06} of the extrasolar planets within 200\,pc of the Earth.}, around 40 are 
known to orbit a component of a binary system \citep{Desidera07}. Indeed, 
\citet{Bonavita07} suggest that after incompleteness has been taken into 
account, the numbers of planets orbiting binary and single stars may be equal.

Given the significant fraction of extrasolar planets orbiting the components of 
binary systems, the dynamical history of such stellar systems becomes 
important for understanding the properties of the planets therein. The 
Kozai mechanism \citep{Kozai62} has been shown by several authors, (e.g. 
\citealt{Innanen97} -- see also \citealt*{Holman97,Takeda05,Malmberg07a}) to 
disrupt the orbits of planets \citep[see e.g.][their fig. 1]{Malmberg07a} that 
form in binary systems.

In this paper we address the possibility of whether the dynamical evolution of 
a cluster may provide a way of inducing the Kozai mechanism in a significant 
fraction of binary stars. We describe the Kozai mechanism in Section~\ref{Kozai}, 
we outline our method in Section~\ref{method}, we present our results in 
Section~\ref{results} and we conclude in Section~\ref{conclusion}. 

\section{The Kozai Mechanism}
\label{Kozai}

The Kozai mechanism \citep{Kozai62} was used to quantify how the orbits of 
inclined asteroids were influenced by Jupiter. It assumes that the mass of the asteroid 
is negligible compared to that of Jupiter and the Sun -- the same assumption can 
be made for a planet orbiting one of the components in a binary system \citep{Innanen97}. 

If the inclination angle of the orbit exceeds 39.23$^\circ$, then the Kozai mechanism states 
that there is a cyclical exchange of angular momentum to the asteroid, causing the eccentricity 
of the asteroid to vary periodically. The same effect is predicted for planetary systems (and  
protoplanetary disks) in binary systems. 

It should be noted that these Kozai cycles can be switched off if the inclination angle 
of the orbit exceeds 140.77$^\circ$. We define a range of inclination angles of systems 
susceptible to the Kozai mechanism as
\begin{equation}
39.23^\circ < i_{\rm Koz} < 140.77^\circ.
\end{equation}

\citet{Malmberg07a} used the Kozai mechanism as a method to disrupt planetary 
systems and showed that it induces a highly eccentric orbit for the outer 
planet, which then crosses the orbits of the inner planets. In some cases this process leads to 
the ejection of one or more of the planets. \citet{Malmberg07a} use intermediate 
($a \sim$100\,AU) separation binary systems to invoke this mechanism.

In our simulations, we investigate whether the inclination angle between the 
components of binary systems can be increased from zero (at the birth 
of the cluster) to an angle in the range $i_{\rm Koz}$ (the \emph{Kozai angle}). 
We assume that the binaries at the cluster birth act as gyroscopes and hence 
any systems with an angle greater than the Kozai angle have undergone 
significant dynamical processing and could be subjected to intense perturbations.

\section{Method}
\label{method}

\subsection{Cluster set-up}

We closely follow the method described by \citet{Parker09a} to set up 
the clusters and binary systems in our simulations. The clusters are designed 
to mimic a `typical' star cluster, similar to Orion with  a mass 
$\sim 10^3$\,M$_\odot$.

For each set of initial conditions, we create a suite of 20 clusters, identical 
apart from the random number seed used to initialise the simulations.

We set our clusters up as initially virialised Plummer spheres 
\citep{Plummer11} using the prescription given in \citet*{Aarseth74}. The 
Plummer sphere provides the positions and velocities of the centre of mass 
of stellar systems. We adopt three different half-mass radii for our clusters; 0.1, 
0.2 and 0.4\,pc. \citet{Parker09a} argue that the initial half-mass radius of 
Orion was in the range 0.1 -- 0.2\,pc, but we include simulations with a half-mass 
radius of 0.4\,pc for comparison.

\subsection{Binary properties}

We create all our clusters with an initial binary fraction, $f_{\rm bin} = 1$ 
(i.e. all stars form in binary systems; there are no singles or triples, 
etc.), where
\begin{equation}
f_{\rm bin} = \frac{B}{S + B},
\end{equation}
and $S$ and $B$ are the numbers of single and binary systems, respectively. 

The mass of the primary star is chosen randomly from a \citet{Kroupa02} IMF of 
the form
\begin{equation} 
 N(M)   \propto 
  \left\{ \begin{array}{ll}
  M^{-1.3} \hspace{0.4cm} m_0 < M/{\rm M_\odot} < m_1   \,, \\
  M^{-2.3} \hspace{0.4cm} m_1 < M/{\rm M_\odot} < m_2   \,,
\end{array} \right.
\end{equation}   
where $m_0$ = 0.1\,M$_\odot$, $m_1$ = 0.5\,M$_\odot$, and 
$m_2$ = 50\,M$_\odot$. For simplicity we do not include brown dwarfs (BDs) 
in our simulations.

Secondary masses are drawn from a flat mass ratio distribution with the 
constraint that if the companion mass is $< 0.1$\,M$_\odot$ it is reselected, 
thereby removing the possibility of choosing secondaries with BD-like masses. 
This constraint maintains the underlying binary fraction, but biases the 
masses of low-mass systems towards unity \citep[see][]{Kouwenhoven09a}.

The generating function for orbital periods are the log-normal
distributions observed by \citet{Duquennoy91} and \citet{Fischer92} of
the form
\begin{equation}
f\left({\rm log}P\right) = C{\rm exp}\left \{ \frac{-{({\rm log}P -
\overline{{\rm log}P})}^2}{2\sigma^2_{{\rm log}P}}\right \},
\end{equation}
where $\overline{{\rm log}P} = 4.8$, $\sigma_{{\rm log}P} = 2.3$ and
$P$ is in days. 

Eccentricities of binary stars are drawn from a thermal eccentricity
distribution \citep{Kroupa95a,Kroupa08} of the form
\begin{equation}
f_e(e) = 2e.
\end{equation}

Binaries with small periods but large eccentricities would 
expect to undergo the tidal circularisation shown in the sample of G-dwarfs 
in \citet{Duquennoy91}. We account for this by reselecting the eccentricity 
if it exceeds the following period-dependent value $e_{\rm tid}$:
\begin{equation}
e_{\rm tid} = \frac{1}{2}\left(0.95 + {\rm tanh}\left(0.6P - 1.7\right)\right).
\end{equation}
This ensures that the eccentricity--period distribution matches the 
observations of \citet{Duquennoy91}, as we expect that tidal circularisation 
occurs before cluster evolution takes place \citep{Parker09b}. Finally, the 
periods are converted to semi-major axes.

By combining the primary and secondary masses of the binaries with
their  semi-major axes and eccentricities, the relative velocity and
radial  components of the stars in each system are determined. These
are then  placed at the centre of mass and centre of velocity for each
system in the  Plummer sphere.

Simulations are run using the \texttt{kira} integrator in Starlab
\citep[e.g.][and references therein]{Zwart99,Zwart01} and evolved  for
10\,Myr.  
 
\subsection{Finding susceptible systems}

For each cluster we identify the binary systems\footnote{We use the 
nearest-neighbour algorithm described by \citet{Parker09a} 
\citep[and independently verified by][]{Kouwenhoven09b} to determine whether 
a star is in a bound binary system.} that have been preserved since the 
birth of the cluster (i.e. those that have not been broken up by dynamical 
processing). We also identify the binary systems formed through dynamical interactions 
in the clusters as these systems are likely to be susceptible to the Kozai mechanism 
\citep{Malmberg07b}.

Simulations show that disk fragmentation is probably a major mode of binary formation 
(\citealp*[e.g.][]{Goodwin04}; \citealp{Goodwin07}) and hence we expect that 
the orbits of the companion and the planets/disk will be roughly coplanar (observations 
of circumstellar disks in binary systems have shown that they are usually inclined by less 
than $10-20^\circ$ to the component stars, e.g. \citealp{Jensen04,Monin06}).  However, 
this is not an assumption that affects our result as we examine the shift in the binary 
inclination relative to whatever initial inclination the planets/disk have.

We measure the change in the inclination of the binary orbit to any planetary system or protoplanetary disk
by examining the change in the orbital angular momentum vector from formation.  We assume that the
inclination of the planets/disk will not be strongly effected by any interaction that changes the inclination
of the binary and hence any change in the inclination of the binary orbit will be a change relative to the
planets/disk.

For the preserved systems we use any change in the angular momentum vector of the binary to 
ascertain whether the system has an inclination angle in the range $i_{\rm Koz}$ after each 
Myr (assuming a birth inclination angle of zero). For the systems formed during the cluster's 
lifetime, we calculate the change in angular momentum with respect to the parental 
binary or binaries\footnote{We define a parental binary as a system that was a binary initially 
 and contains one of the stars in a newly formed binary. The \citet{Duquennoy91} 
period generating function leads to very wide (unphysical) binaries in the clusters and so not 
all the primordial binaries created in the simulations are detected by our algorithm 
\citep[see][for a more detailed discussion]{Parker09a}. The result of this is that some of the 
newly formed binaries may only have one parental binary.} and use the largest angle in our 
calculations\footnote{In practice, in the very few binaries created through these intense 
dynamical interactions, both angles tend to be in the range $i_{\rm Koz}$ so the angle chosen is irrelevant.}.

As we are interested in the effect of cluster evolution on planetary systems 
(or protoplanetary disks), we define a periastron distance that allows for the 
formation of such a system without interference from the secondary component of the 
binary system. We take the standard definition of periastron distance $r_{\rm peri}$:
\begin{equation}
r_{\rm peri} = a(1 - e),
\end{equation}
where $a$ and $e$ are the semi-major axis and the eccentricity of the binary system, respectively.

We determine  $r_{\rm peri}$ for each system, and assume that a stable planetary system (or disk) 
could have developed for any binary with $r_{\rm peri} > 100$\,AU. Of the 43 planets orbiting 
a binary component in the sample of \citet{Desidera07}, only 5 are in binary systems with separations 
less than 100\,AU. If a binary exceeds this periastron threshold we determine its inclination angle 
from the change in the orbital angular momentum vector.

\section{Results}
\label{results}

\subsection{The fraction of systems susceptible to the Kozai mechanism}

\begin{figure}
  \begin{center}
\rotatebox{270}{\includegraphics[scale=0.4]{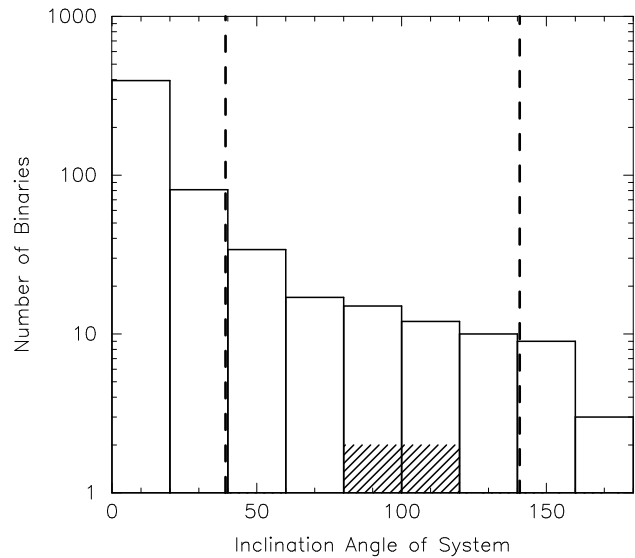}}
  \end{center}
  \caption[bf]{The distribution of inclination angles of systems with $r_{\rm peri} > 100$\,AU 
after 1\,Myr for clusters with an initial half-mass radius of 0.1\,pc. Binary systems from 
20 different realisations of the same cluster have been binned together in this 
histogram. The open histogram represents primordial binaries, preserved since the birth of 
the cluster; the hashed histogram represents binary systems formed during the dynamical evolution 
of the cluster. The dashed lines indicate the threshold angles (39.23$^\circ$ and 140.77$^\circ$), between 
which the Kozai mechanism may act upon any planets in orbit around the binary stars.}
  \label{theta_dist}
\end{figure}

We show the distribution of inclination angles for systems with $r_{\rm peri} > 100$\,AU after 
1\,Myr in Fig.~\ref{theta_dist}. Whilst only 1~per~cent of \emph{all} systems have an inclination 
angle within the threshold range (39.23$^\circ$ to 140.77$^\circ$), many of the binaries are tight 
($<$\,30\,AU) and hence not susceptible to dynamical processing \citep{Parker09a}. 

\begin{figure}
  \begin{center}
\rotatebox{270}{\includegraphics[scale=0.4]{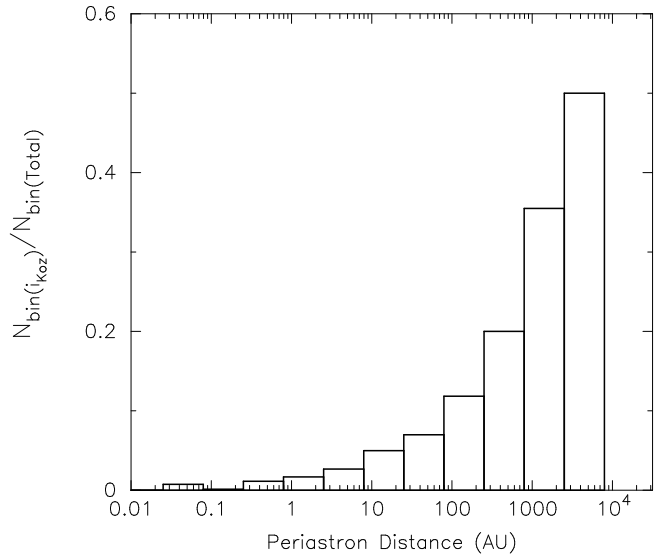}}
  \end{center}
  \caption[bf]{The fraction of binaries in the range $i_{\rm Koz}$ as a function of perisatron 
distance $r_{\rm peri}$ after 1\,Myr of cluster evolution (initial half-mass radius, 
$r_{\rm 1/2}=0.1$\,pc).}
  \label{peri_kozai}
\end{figure}

This is readily demonstrated in Fig.~\ref{peri_kozai}, where we plot the fraction of binaries 
with inclination angles in the threshold range $i_{\rm Koz}$ as a function of $r_{\rm peri}$. Binaries 
with $r_{\rm peri} > 100$\,AU are far more likely to have inclination angles in the range $i_{\rm Koz}$ 
than the tighter systems. Additionally, a binary with a periastron distance $r_{\rm peri} < 100$\,AU is unlikely 
to form a stable planetary system, as that system will be perturbed by the other component of the binary system.

The contribution from binaries that are formed during the cluster's evolution is minimal. As can be 
seen in Fig.~\ref{theta_dist}, only four newly formed systems with $r_{\rm peri} > 100$\,AU lie 
within the threshold range $i_{\rm Koz}$, compared to the many tens of primordial binaries.

\begin{figure}
  \begin{center}
\rotatebox{270}{\includegraphics[scale=0.4]{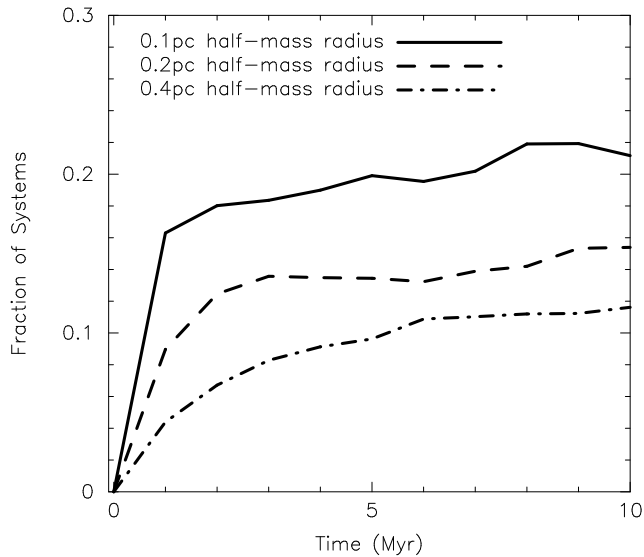}}
  \end{center}
  \caption[bf]{The fraction of systems with $r_{\rm peri} > 100$\,AU and 
an inclination angle within the Kozai threshold range $i_{\rm Koz}$ 
(39.23$^\circ$ -- 140.77$^\circ$) as a function of time. We show the results for 
three different initial cluster half-mass radii; 0.1\,pc, 0.2 \,pc and 
0.4\,pc. For the most dense cluster, the fraction of systems that could be subjected 
to the Kozai mechanism remains at $\sim$ 20 per cent during the first 10\,Myr of cluster evolution.}
  \label{kozai_frac}
\end{figure}

Of the systems that have $r_{\rm peri} > 100$\,AU, we determine the fraction of systems that 
have an inclination angle between 39.23$^\circ$ and 140.77$^\circ$. We show the evolution of 
this fraction during the cluster's lifetime in Fig.~\ref{kozai_frac}. We show the fraction 
of systems that could be subjected to the Kozai mechanism for three different initial 
half-mass radii; 0.1\,pc (the solid line), 0.2\,pc (the dashed line) and 0.4\,pc (the dashed-dot line). 

For the most dense clusters ($r_{1/2} = 0.1$ pc), the fraction of systems that could 
potentially undergo the Kozai mechanism is roughly constant, at $\sim$ 20 per cent. For 
the less dense clusters, the fraction is less; $\sim$ 13 per cent for $r_{1/2} = 0.2$ pc, 
and $\sim$ 10 per cent for $r_{1/2} = 0.4$ pc. This decrease in affected systems with 
increasing half-mass radius is simply due to there being fewer interactions in the 
less-dense clusters.

The fraction of systems that could be subjected to the Kozai mechanism remains roughly 
constant after 1\,Myr for the most dense clusters ($r_{1/2} = 0.1$ \& $0.2$ pc). This is entirely 
due to the crossing time of the cluster being only $\sim 0.2$ Myr \citep{Parker09a} and therefore 
the cluster has relaxed and the majority of binaries have already reached dynamical equilibrium.

\subsection{The Kozai timescale}

A planet orbiting the component of a binary star will undergo Kozai cycles on the following 
timescale \citep*{Kiseleva98,Takeda08,Verrier09}:  
\begin{equation}
\tau_{\rm Koz} \simeq \frac{2}{3\pi} \frac{P_{\rm bin}^2}{P_p}(1 - e_{\rm bin}^2)^{3/2}\frac{m_1 + m_2 + m_p}{m_2},
\label{koztime}
\end{equation}
where $P_{\rm bin}$ is the period of the binary, $P_p$ is the period of the planet, $e_{\rm bin}$ is the 
eccentricity of the binary, $m_1$ and $m_2$ are the masses of the primary and secondary components of the 
binary respectively, and $m_p$ is the mass of the planet.

For the densest clusters ($r_{1/2} = 0.1$\,pc), we use Eqn.~\ref{koztime} to determine the Kozai timescale 
for two hypothetical planets orbiting a component of each binary. One planet has a period and mass similar to 
that of Neptune (165\,years and $5\times10^{-5}$M$_\odot$ respectively), the other has a period and mass 
similar to Jupiter (12\,years and $9\times10^{-4}$M$_\odot$ respectively). 

\begin{figure}
  \begin{center}
\rotatebox{270}{\includegraphics[scale=0.4]{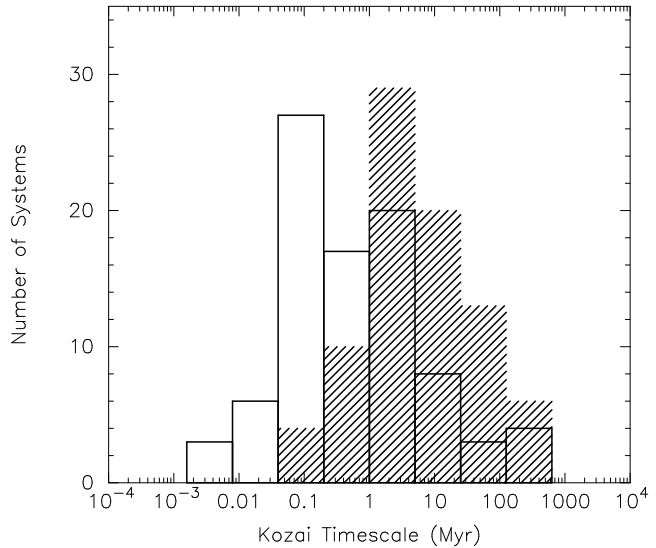}}
  \end{center}
  \caption[bf]{The distribution of Kozai timescales, $\tau_{\rm Koz}$, in Myr, for two hypothetical planets. The open 
histogram is the distribution for a planet with Neptune's mass ($5\times10^{-5}$M$_\odot$) and period (165\,years), and 
the hashed histogram is the distribution for a planet with Jupiter's mass ($9\times10^{-4}$M$_\odot$) and period (12\,years).}
  \label{kozai_time}
\end{figure}

Our results are shown in Fig.~\ref{kozai_time}. We determine the Kozai timescale for the hypothetical planets for 
all binaries with a periastron distance $r_{\rm peri} > 100$\,AU and an inclination angle $>$ 39.23$^\circ$. 
We show the distribution of Kozai timescales for the Neptune analog in the open histogram, and the Kozai timescales 
for the Jupiter analog in the hatched histogram. 

For the Neptune analog, there are a large number of systems in which the Kozai timescale is much less than 1\,Myr. This 
means that such planets could be subjected to the Kozai mechanism very early on in the dynamical evolution of the 
cluster, and given this timescale it is more pertinent to ascribe these effects to a protoplanetary disk rather than a 
system of fully formed planets. It should also be noted that Kozai cycles could occur several times in 1\,Myr for each 
planetary system. 

The Kozai timescale is longer for the Jupiter analog, and we would expect on average only one Kozai cycle per Myr for 
such a planet. However, if a system consists of more than one planet, it is the timescale for the outer planet in the 
system that is important \citep{Takeda08}. Therefore a Jupiter-like planet could undergo more than one Kozai cycle per 
Myr if there were also planets orbiting the star with longer periods.

\section{Conclusions}
\label{conclusion}

We use $N$-body simulations to dynamically evolve a typical Orion-like star cluster for 
10\,Myr. We examine the effect of dynamical evolution on binary systems that could host 
a stable planetary system or protoplanetary disk. To this end we determine the fraction 
of systems with a periastron distance $r_{\rm peri} > 100$\,AU where the secondary 
component of the binary is shifted with respect to the primary by an angle within the Kozai 
threshold range (39.23$^\circ < i_{\rm Koz} < $ 140.77$^\circ$).

For typical clusters with an initial half-mass radius corresponding to that of  Orion originally 
\citep[0.1\, -- 0.2\,pc; see][for a discussion of the primordial half-mass radius of Orion]{Parker09a}, 
we find that 20 per cent of binary systems with $r_{\rm peri} > 100$\,AU 
have inclination angles in the range $i_{\rm Koz}$ that could induce the Kozai mechanism 
\citep{Kozai62}.  The Kozai mechanism has been shown to drastically alter 
the orbital properties of planets \citep{Innanen97,Malmberg07a} and in particular the eccentricities 
of extrasolar planets \citep{Malmberg09}.

In these dense clusters, we place two hypothetical planets in systems that could be subjected to 
the Kozai mechanism in order to determine the Kozai timescale, i.e. the length of time it takes for 
a planet to undergo Kozai cycles. For Neptune-like planets, we find that the Kozai mechanism has 
the potential to occur more than once a Myr, with a longer timescale for Jupiter-like planets.  

For less dense clusters, the fraction of planetary systems in $> 100$\,AU binaries that could be subjected to 
the Kozai mechanism is still between 10 and 20 per cent, i.e. a not insignificant fraction.

Around a quarter of the extrasolar planets discovered to date are orbiting a component of a binary 
system with a separation exceeding 100\,AU \citep{Desidera07}. Due to incompleteness in these observations, 
it is suggested that half of all extrasolar planets may be in binary systems \citep{Bonavita07}. We 
 therefore propose that simple dynamical processing of binary stars in clusters could feasibly 
affect up to 10 per cent of \emph{all} extrasolar planetary systems by inducing the Kozai 
mechanism.

\section*{Acknowledgements}

We thank Melvyn Davies and Daniel Malmberg for useful discussions and comments on the original text. 
RJP acknowledges financial support from STFC. This work has made use of the Iceberg computing 
facility, part of the White Rose Grid computing facilities at the University of Sheffield.

\bibliographystyle{mn2e}
\bibliography{Koz_ref}

\label{lastpage}

\end{document}